\documentclass[aps,singlecolumn,superscriptaddress,superscriptaddress,amsthm,amsmath,amssymb]{revtex4-1}  
\usepackage{graphicx}
\usepackage{dcolumn}
\usepackage{bm}
\usepackage{hyperref}
\usepackage{mathtools}
\usepackage{color}
\usepackage{xcolor}
\usepackage{adjustbox}
\usepackage[normalem]{ulem}
\usepackage{caption}

\usepackage{xy}
\xyoption{matrix}
\xyoption{frame}
\xyoption{arrow}
\xyoption{arc}

\usepackage{ifpdf}
\ifpdf
\else
\PackageWarningNoLine{Qcircuit}{Qcircuit is loading in Postscript mode.  The Xy-pic options ps and dvips will be loaded.  If you wish to use other Postscript drivers for Xy-pic, you must modify the code in Qcircuit.tex}
\xyoption{ps}
\xyoption{dvips}
\fi

\entrymodifiers={!C\entrybox}

\newcommand{\bra}[1]{{\left\langle{#1}\right\vert}}
\newcommand{\ket}[1]{{\left\vert{#1}\right\rangle}}
\newcommand{\qw}[1][-1]{\ar @{-} [0,#1]}
\newcommand{\qwx}[1][-1]{\ar @{-} [#1,0]}

\newcommand{\gate}[1]{*+<.6em>{#1} \POS ="i","i"+UR;"i"+UL **\dir{-};"i"+DL **\dir{-};"i"+DR **\dir{-};"i"+UR **\dir{-},"i" \qw}

\newcommand{\control}{*!<0em,.025em>-=-<.2em>{\bullet}}

\newcommand{\ctrl}[1]{\control \qwx[#1] \qw}

\newcommand{\targ}{*+<.02em,.02em>{\xy ="i","i"-<.39em,0em>;"i"+<.39em,0em> **\dir{-}, "i"-<0em,.39em>;"i"+<0em,.39em> **\dir{-},"i"*\xycircle<.4em>{} \endxy} \qw}

\newcommand{\multigate}[2]{*+<1em,.9em>{\hphantom{#2}} \POS [0,0]="i",[0,0].[#1,0]="e",!C *{#2},"e"+UR;"e"+UL **\dir{-};"e"+DL **\dir{-};"e"+DR **\dir{-};"e"+UR **\dir{-},"i" \qw}
\newcommand{\ghost}[1]{*+<1em,.9em>{\hphantom{#1}} \qw}
\newcommand{\Qcircuit}{\xymatrix @*=<0em>}

\def\tableone{
	\begin{tabular}{|c|c|c|c|c|c|c|c|c|c|c|}
		\hline
		\multicolumn{1}{|c|}{Graph}
		&\multicolumn{4}{c|}{Pre-Fault Tolerant}
		&\multicolumn{3}{c|}{Fault-Tolerant (Gridsynth)}
		&\multicolumn{3}{c|}{Fault-Tolerant (RUS)} \\ \cline{2-11}
		& $n$ & $\cnotgate$ & $\rzgate$ & Depth & $n$ & $\cnotgate$ & $\tgate$ & $n$ & $\cnotgate$ & $\tgate$ 
		\\ \hline \hline
		\multicolumn{11}{|c|}{4th order} \\ \cline{1-11}
		$(3{-}5{-}70)$ & 70 & 648,885 & 700,367 & 25,333 & 126 & 4,751,775 & 13,398,840 & 131 & 9,573,795 & 6,751,395 \\ \hline
		$(4{-}4{-}98)$ & 98 & 1,590,750 & 1,696,926 & 41,811 & 179 & 11,041,734 & 23,878,140 & 185 & 19,079,280 & 12,615,702 \\ \hline
		$(5{-}3{-}72)$ & 72 & 1,924,560 & 2,007,822 & 66,525 & 142 & 13,568,294 & 28,047,884 & 148 & 22,865,870 & 15,011,834 \\ \hline \hline
		\multicolumn{11}{|c|}{6th order} \\ \cline{1-11}
		$(3{-}5{-}70)$ & 70 & 1,873,350 & 2,021,797 & 72,587 & 126 & 9,802,780 & 26,068,380 & 131 & 18,995,860 & 13,210,730 \\ \hline
		$(4{-}4{-}98)$ & 98 & 6,011,250 & 6,412,126 & 155,259 & 179 & 37,283,365 & 77,435,190 & 185 & 63,102,200 & 41,364,575 \\ \hline
		$(5{-}3{-}72)$ & 72 & 4,519,800 & 4,715,202 & 154,005 & 142 & 24,882,660 & 49,081,320 & 148 & 40,935,060 & 26,578,440 \\ \hline
		$(7{-}2{-}50)$ & 50 & 1,124,700 & 1,160,419 & 51,795 & 96 & 6,991,712 & 12,887,472 & 101 & 11,154,260 & 7,172,048 \\ \hline
	\end{tabular}
}

\def\tabletwo{
	\begin{tabular}{|c|c|c|c|c|c|c|c|c|c|} 
		\hline
		\multicolumn{1}{|c|}{Graph ($k{-}d{-}n$)}
		&\multicolumn{3}{c|}{Pre-fault tolerant cost}
		&\multicolumn{3}{c|}{Fault-tolerant cost (Gridsynth)}
		&\multicolumn{3}{c|}{Fault-tolerant cost (RUS)} \\ \cline{2-10}
		& $n$ & $\cnotgate$ & $\rzgate$ & $n$ & $\cnotgate$ & $\tgate$ & $n$ & $\cnotgate$ & $\tgate$
		\\ \hline \hline 
		$(3{-}1{-}4)$ & 4 & 21,420 & 22,854 & 6 & 44,546 & 656,999 & 6 & 312,218 & 288,084 \\ \hline 
		$(3{-}d{-}5)$ & 5 & 25,740 & 28,606 & 7 & 56,538 & 1,092,171 & 7 & 404,791 & 373,904 \\ \hline 
		$(3{-}d{-}6)$ & 6 & 42,030 & 45,774 & 10 & 105,944 & 1,354,329 & 12 & 654,011 & 595,619 \\ \hline 
		$(3{-}d{-}7)$ & 7 & 34,527 & 37,921 & 11 & 109,184 & 1,053,768 & 13 & 533,148 & 470,965 \\ \hline 
		$(3{-}d{-}8)$ & 8 & 49,014 & 52,915 & 15 & 190,432 & 1,402,834 & 18 & 743,399 & 640,570 \\ \hline 
		$(3{-}d{-}9)$ & 9 & 61,776 & 67,404 & 16 & 272,375 & 1,874,968 & 19 & 1,003,682 & 865,932 \\ \hline 
		$(3{-}d{-}10)$ & 10 & 65,850 & 71,132 & 18 & 282,579 & 1,970,933 & 21 & 1,053,528 & 906,787  \\ \hline 
		$(3{-}d{-}11)$ & 11 & 68,364 & 74,287 & 19 & 297,681 & 1,950,765 & 22 & 1,058,286 & 897,899 \\ \hline 
		$(3{-}d{-}12)$ & 12 & 81,840 & 88,896 & 22 & 379,351 & 2,151,939 & 25 & 1,213,604 & 1,001,577 \\ \hline 
		\hline
		$(4{-}1{-}5)$ & 5 & 31,752 & 33,523 & 7 & 70,338 & 973,842 & 7 & 464,312 & 428,246 \\ \hline 
		$(4{-}d{-}6)$ & 6 & 33,894 & 36,054 & 10 & 101,048 & 917,418 & 12 & 466,553 & 411,243 \\ \hline 
		$(4{-}d{-}7)$ & 7 & 47,475 & 50,649 & 11 & 174,643 & 1,154,595 & 13 & 626,252 & 528,012 \\ \hline 
		$(4{-}d{-}8)$ & 8 & 56,952 & 60,353 & 15 & 231,522 & 1,493,720 & 18 & 812,052 & 690,186 \\ \hline 
		$(4{-}d{-}9)$ & 9 & 70,848 & 75,288 & 16 & 313,316 & 1,832,632 & 19 & 1,019,675 & 857,497 \\ \hline 
		$(4{-}d{-}10)$ & 10 & 74,445 & 78,713 & 18 & 330,598 & 1,928,493 & 21 & 1,074,228 & 902,955 \\ \hline 
		$(4{-}d{-}11)$ & 11 & 88,101 & 93,387 & 19 & 399,603 & 2,128,496 & 22 & 1,217,455 & 1,000,280  \\ \hline 
		$(4{-}d{-}12)$ & 12 & 89,964 & 95,692 & 22 & 434,945 & 2,220,409 & 25 & 1,282,712 & 1,052,963 \\ \hline 
		\hline
		$(5{-}1{-}6)$ & 6 & 37,584 & 39,448 & 10 & 119,778 & 930,247 & 12 & 486,559 & 421,418 \\ \hline 
		$(5{-}d{-}7)$ & 7 & 37,758 & 39,797 & 11 & 143,125 & 866,847 & 13 & 479,031 & 399,411 \\ \hline 
		$(5{-}d{-}8)$ & 8 & 66,636 & 69,732 & 15 & 275,838 & 1,627,403 & 18 & 902,075 & 760,854 \\ \hline 
		$(5{-}d{-}9)$ & 9 & 62,520 & 65,658 & 16 & 269,112 & 1,551,995 & 19 & 861,283 & 730,606 \\ \hline 
		$(5{-}d{-}10)$ & 10 & 87,354 & 91,166 & 18 & 397,923 & 2,036,766 & 21 & 1,174,295 & 966,111  \\ \hline 
		$(5{-}d{-}11)$ & 11 & 89,817 & 94,109 & 19 & 401,968 & 2,179,369 & 22 & 1,234,464 & 1,029,598 \\ \hline 
		$(5{-}d{-}12)$ & 12 & 109,836 & 115,276 & 22 & 541,352 & 2,445,309 & 25 & 1,462,680 & 1,177,035 \\ \hline 
		\hline
		$(7{-}1{-}8)$ & 8 & 76,440 & 78,412 & 15 & 362,435 & 1,814,698 & 18 & 1,046,652 & 873,457 \\ \hline 
		$(7{-}d{-}9)$ & 9 & 63,510 & 65,712 & 16 & 296,818 & 1,482,322 & 19 & 854,973 & 711,118 \\ \hline 
		$(7{-}d{-}10)$ & 10 & 72,072 & 74,633 & 18 & 340,054 & 1,494,097 & 21 & 899,034 & 723,474 \\ \hline 
		$(7{-}d{-}11)$ & 11 & 83,283 & 85,645 & 19 & 402,510 & 1,729,234 & 22 & 1,048,114 & 838,359 \\ \hline 
		$(7{-}d{-}12)$ & 12 & 105,534 & 109,158 & 22 & 539,611 & 2,026,914 & 25 & 1,288,877 & 996,683  \\ \hline 
	\end{tabular}
}

\def\tablethree{
	\begin{tabular}{|c|c|c|c|c|c|c|c|c|c|} 
		\hline
		\multicolumn{1}{|c|}{Graph ($k{-}d{-}n$)}
		&\multicolumn{3}{c|}{Pre-fault tolerant cost}
		&\multicolumn{3}{c|}{Fault-tolerant cost (Gridsynth)}
		&\multicolumn{3}{c|}{Fault-tolerant cost (RUS)} \\ \cline{2-10}
		& $n$ & $\cnotgate$ & $\rzgate$ & $n$ & $\cnotgate$ & $\tgate$ & $n$ & $\cnotgate$ & $\tgate$
		\\ \hline \hline 
		$(3{-}1{-}4)$ & 4 & 19,050 & 20,326 & 6 & 40,104 & 559,199 & 6 & 267,051 & 245,811 \\ \hline 
		$(3{-}d{-}5)$ & 5 & 27,900 & 31,006 & 7 & 59,300 & 851,919 & 7 & 403,992 & 372,413 \\ \hline 
		$(3{-}d{-}6)$ & 6 & 43,200 & 47,048 & 10 & 105,779 & 1,265,212 & 12 & 615,321 & 557,897 \\ \hline 
		$(3{-}d{-}7)$ & 7 & 44,370 & 48,729 & 11 & 127,795 & 1,162,152 & 13 & 591,622 & 519,905 \\ \hline 
		$(3{-}d{-}8)$ & 8 & 61,740 & 66,651 & 15 & 216,250 & 1,512,291 & 18 & 808,460 & 693,154 \\ \hline 
		$(3{-}d{-}9)$ & 9 & 73,260 & 79,932 & 16 & 292,799 & 1,921,586 & 19 & 1,039,192 & 892,361 \\ \hline 
		$(3{-}d{-}10)$ & 10 & 87,000 & 93,974 & 18 & 335,036 & 2,227,206 & 21 & 1,202,108 & 1,030,323 \\ \hline 
		$(3{-}d{-}11)$ & 11 & 98,415 & 106,935 & 19 & 382,947 & 2,408,245 & 22 & 1,320,426 & 1,116,307 \\ \hline 
		$(3{-}d{-}12)$ & 12 & 114,855 & 124,751 & 22 & 486,069 & 2,625,050 & 25 & 1,501,123 & 1,231,756 \\ \hline 
		\hline
		$(4{-}1{-}5)$ & 5 & 30,780 & 32,497 & 7 & 64,205 & 842,883 & 7 & 404,922 & 372,437 \\ \hline 
		$(4{-}d{-}6)$ & 6 & 44,730 & 47,578 & 10 & 117,715 & 1,051,819 & 12 & 534,980 & 471,445 \\ \hline 
		$(4{-}d{-}7)$ & 7 & 57,375 & 61,209 & 11 & 191,469 & 1,221,648 & 13 & 667,424 & 560,700 \\ \hline 
		$(4{-}d{-}8)$ & 8 & 69,720 & 73,881 & 15 & 264,421 & 1,655,404 & 18 & 906,203 & 768,382 \\ \hline 
		$(4{-}d{-}9)$ & 9 & 93,120 & 98,952 & 16 & 359,050 & 2,011,995 & 19 & 1,130,662 & 946,685 \\ \hline 
		$(4{-}d{-}10)$ & 10 & 107,100 & 113,234 & 18 & 428,640 & 2,415,220 & 21 & 1,357,074 & 1,136,837 \\ \hline 
		$(4{-}d{-}11)$ & 11 & 130,455 & 138,275 & 19 & 534,838 & 2,764,991 & 22 & 1,595,466 & 1,306,936 \\ \hline 
		$(4{-}d{-}12)$ & 12 & 141,120 & 150,096 & 22 & 623,051 & 3,096,764 & 25 & 1,803,268 & 1,477,260 \\ \hline 
		\hline
		$(5{-}1{-}6)$ & 6 & 44,550 & 46,758 & 10 & 129,405 & 982,123 & 12 & 516,259 & 446,710 \\ \hline 
		$(5{-}d{-}7)$ & 7 & 54,405 & 57,339 & 11 & 187,022 & 1,111,146 & 13 & 618,576 & 515,215 \\ \hline 
		$(5{-}d{-}8)$ & 8 & 72,360 & 75,721 & 15 & 294,048 & 1,684,602 & 18 & 939,612 & 790,538 \\ \hline 
		$(5{-}d{-}9)$ & 9 & 82,800 & 86,952 & 16 & 328,222 & 1,841,418 & 19 & 1,029,388 & 870,758 \\ \hline 
		$(5{-}d{-}10)$ & 10 & 108,330 & 113,054 & 19 & 446,178 & 2,218,756 & 21 & 1,289,821 & 1,057,906 \\ \hline 
		$(5{-}d{-}11)$ & 11 & 132,000 & 138,615 & 20 & 521,262 & 2,794,246 & 22 & 1,591,495 & 1,327,663 \\ \hline 
		$(5{-}d{-}12)$ & 12 & 157,140 & 164,916 & 22 & 699,399 & 3,077,140 & 25 & 1,858,714 & 1,491,134 \\ \hline 
		\hline
		$(7{-}1{-}8)$ & 8 & 76,440 & 78,412 & 15 & 332,706 & 1,607,197 & 18 & 935,256 & 776,724 \\ \hline 
		$(7{-}d{-}9)$ & 9 & 89,610 & 92,712 & 16 & 365,610 & 1,781,648 & 19 & 1,035,870 & 859,263 \\ \hline 
		$(7{-}d{-}10)$ & 10 & 99,990 & 103,539 & 19 & 436,370 & 1,879,136 & 21 & 1,135,877 & 911,157 \\ \hline 
		$(7{-}d{-}11)$ & 11 & 115,020 & 118,276 & 20 & 507,581 & 2,177,549 & 22 & 1,321,786 & 1,058,741 \\ \hline 
		$(7{-}d{-}12)$ & 12 & 141,570 & 146,426 & 22 & 646,853 & 2,360,222 & 25 & 1,513,669 & 1,165,082 \\ \hline 
	\end{tabular}
}

\begin{document}


\global\long\def\U{\mathbb{U}}
\global\long\def\l({\left(}
\global\long\def\r){\right)}


\global\long\def\lc{\left\{}
\global\long\def\rc{\right\}}
\global\long\def\R#1{R_{\left|#1\right\rangle }}
\global\long\def\bra#1{\left\langle #1\right|}
\global\long\def\ket#1{\left|#1\right\rangle }
\global\long\def\op#1#2{\left|#1\right\rangle \left\langle #2\right|}
\global\long\def\ip#1#2{\left\langle #1,#2\right\rangle }
\global\long\def\s#1{\sum_{#1\in\left\{  0,1\right\}  ^{k}}}
\global\long\def\I{\mathbf{I}}
\global\long\def\ve{\varepsilon}

\newcommand{\eq}[1]{(\ref{eq:#1})}
\renewcommand{\sec}[1]{\hyperref[sec:#1]{Section~\ref*{sec:#1}}}
\newcommand{\fig}[1]{\hyperref[fig:#1]{Figure~\ref*{fig:#1}}}
\newcommand{\tab}[1]{\hyperref[tab:#1]{Table~\ref*{tab:#1}}}
\newcommand{\routine}[1]{\hyperref[#1]{Routine~\ref*{#1}}}

\newcommand{\Gate}[1]{\textsc{#1}}
\newcommand{\hgate}{\Gate{h}}
\newcommand{\zgate}{\Gate{z}}
\newcommand{\czgate}{\Gate{cz}}
\newcommand{\sgate}{\Gate{s}}
\newcommand{\tgate}{\Gate{t}}
\newcommand{\pgate}{\Gate{p}}
\newcommand{\notgate}{\Gate{not}}
\newcommand{\cnotgate}{\Gate{cnot}}
\newcommand{\swapgate}{\Gate{swap}}
\newcommand{\rzgate}{{\Gate{r}}_{z}}
\newcommand{\rygate}{{\Gate{r}}_{y}}
\newcommand{\rxgate}{{\Gate{r}}_{x}}

\newcommand{\GF}{\operatorname{GF}}
\DeclarePairedDelimiter\norm{\|}{\|}

\newcommand{\new}[1]{{\color{black}{#1}}}
\newcommand{\old}[1]{}

\newtheorem{lemma}{Lemma}

\title{Low cost quantum circuits for classically intractable instances of the \\ Hamiltonian dynamics simulation problem}

\author{Yunseong Nam}
\email{nam@ionq.co}
\affiliation{IonQ, College Park, MD 20740, USA}
\author{Dmitri Maslov}
\email{dmitri.maslov@gmail.com}
\affiliation{National Science Foundation, Alexandria, VA 22314, USA}
\altaffiliation{now with IBM Thomas J. Watson Research Center, Yorktown Heights, NY 10598, USA}

%
%
%

\date{\today}

\begin{abstract}
We develop circuit implementations for digital-level quantum Hamiltonian dynamics simulation algorithms suitable for implementation on a reconfigurable quantum computer, such as trapped ions.  Our focus is on the co-design of a problem, its solution, and quantum hardware capable of executing the solution at the minimal cost expressed in terms of the quantum computing resources used, while demonstrating the solution of an instance of a scientifically interesting problem that is intractable classically.  The choice for Hamiltonian dynamics simulation is due to the combination of its usefulness in the study of equilibrium in closed quantum mechanical systems, a low cost in the implementation by quantum algorithms, and the difficulty of classical simulation.  By targeting a specific type of quantum computer and tailoring the problem instance and solution to suit physical constraints imposed by the hardware, we are able to reduce the resource counts by a factor of $10$ in a physical-level implementation and a factor of $30$ to $60$ in a fault-tolerant implementation over state of the art. 
\end{abstract}

\maketitle


\section{Introduction} 

Quantum supremacy is a computational experiment designed to demonstrate a computational capability of a quantum machine that cannot be matched by a classical computer.  It is highly relevant to this paper, since we too focus on a quantum computation of the size that cannot be performed via classical means.  Quantum supremacy is an important milestone in the development of quantum computers.  Multiple IT giants are targeting quantum supremacy \cite{www:companies}, and it is perhaps reasonable to anticipate that a successful demonstration may be obtained within at most a few years. 

Once quantum supremacy is demonstrated, a next step is to go beyond what the supremacy would have achieved.  The supremacy experiment proposed in \cite{ar:bisb}, in particular, reduces to the execution of a random quantum circuit on a quantum computer that is too large for a classical computer to cope with simulating.  In this work, rather than focusing on an artificial problem designed purely for demonstrating quantum supremacy \cite{ar:bisb}, we target the selection of a known computational problem and a specific input instance, such that, to the best of our knowledge, the problem/instance pair requires a classically intractable computation.  We develop an optimized quantum circuit computing the answer for the selected problem/instance pair that can be suitable for the execution on near-term quantum computers.  Our overarching goal is to select a problem/instance pair and develop a short enough circuit such that it will be the smallest among all circuits solving a post-classical instance of a scientific problem.  A quantum computation described by such circuit constitutes a qualitative step forward, where a quantum computer can now be thought of as being a tool in the solution of a problem rather than the focus of the study.  A more advanced demonstration past the one we are reporting in this work could target the solution of a problem with a commercial value.

In the ``Results'' section, we report the quantum resources required for running the proposed post-supremacy experiments and discuss the quality of our results, expressed in terms of quantum circuit gate count and depth, and viewed through the lens of comparisons to prior and similar-spirited work. We summarize and discuss our overall findings in the ``Discussion'' section. In Section ``Methods'', we introduce the details of the problem and the specific instance of this problem that we are proposing as satisfying the conditions outlined in the previous paragraph.  \new{We prove that the type of problem we consider is classically hard under the assumption that Polynomial Hierarchy does not collapse.} We also describe our solution, including numerous techniques used to improve quantum computing resources used. We stress that we develop complete and fully specified quantum circuits as a part of this study.

\section{Results}

\begin{table}[t]
\renewcommand\thetable{1}
\caption{Gate counts for our proposed experiments}
\label{tab:results}
\tableone
\end{table}

\tab{results} reports gate counts in the post-supremacy Hamiltonian dynamics simulation with Heisenberg interactions and random disorders on graphs with small diameter (Readers are strongly encouraged to read section ``Methods'' for technical description of the Hamiltonian and the optimized implementation detail of the simulation, used throughout this section).  We show resource counts targeted for both physical-level and fault-tolerant implementations.  We note that the counts were obtained while optimizing the circuit depth.  We found that they may be improved by about 20\% if we optimize the gate counts themselves instead.  

\begin{figure}[t]
\includegraphics[scale=1.5]{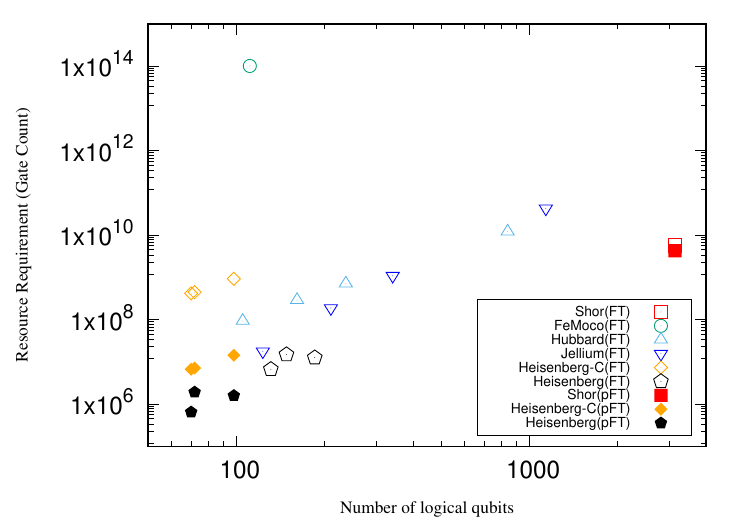}
   \caption{\new{Quantum resource counts as a function of the number of logical qubits for various quantum circuits. Solid plot symbols denote physical-level/pre-fault tolerant (pFT) circuits.  Hollow plot symbols denote fault tolerant (FT) circuits.  For pFT circuits, we use $\cnotgate$ gate counts as the resource requirement and the logical qubit count is the same as the bare physical qubit count.  For FT circuits, we use $\tgate$ gate counts as the resource requirement.  For the various similarly-spirited previous work, see \cite{Kutin} for Shor, \cite{Reiher} for FeMoco, \cite{ar:bgb} for Hubbard and Jellium, and \cite{arXiv:1711.10980} for Heisenberg Hamiltonian over cycle (C).  The 4th order formula results reported in this paper (Table \ref{tab:results}) appear as `Heisenberg' in the figure.}}
   \label{fig:comparison}
\end{figure}

For the purpose of comparison to prior\new{, similarly spirited} work reporting detailed gate counts \new{(see \fig{comparison} for a visual representation of a more comprehensive list of data)}, we focus on our best result, the Hamiltonian simulation over the $(3{-}5{-}70)$ graph (see section ``Methods'' for detail).  In quantum chemistry, the simulation of FeMoco (the primary cofactor of nitrogenase, which is an enzyme used in the nitrogen fixation) required the circuit with $10^{14}$ $\tgate$ gates over $111$ qubits \cite{Reiher}.  In comparison, our fault-tolerant circuits \new{simulating a Heisenberg Hamiltonian system} with $6.8{\times}10^6$ and $1.3{\times}10^7$ $\tgate$ gates (4th and 6th order product formulas, correspondingly) are significantly shorter, while relying on a comparable number of qubits, $131$, and corresponding to solving a task of a similar classical complexity.  To factor a 1,024-digit integer number, \cite{Kutin} constructed a circuit with $5.7{\times}10^9$ $\tgate$ gates spanning 3,132 qubits.  Our circuits \new{simulating a Heisenberg Hamiltonian system} are orders of magnitude shorter and operate over a much smaller number of qubits.  Recent work \cite{ar:bgb} required $10^8$ $\tgate$ gates to solve a problem in the study of solid-state electronic structure, whereas our $\tgate$ count is only $6.8{\times}10^6$.  Finally, the task of a very similar complexity ($70$ qubits, $\varepsilon=0.001$, Hamiltonian on a cycle \cite{arXiv:1711.10980}) is solved using $4.2{\times}10^8$ $\tgate$ gates or $6.7{\times}10^6$ $\cnotgate$ gates, showing the advantage of our approach by a factor of $60$ in the $\tgate$ count and a factor of $10$ in the $\cnotgate$ count.  In fact, the circuits we developed to simulate the Hamiltonian evolution are so short, that we hope they may be possible to execute on pre-fault tolerant quantum computers.  

The two-qubit depth of our quantum circuits is very small, making them particularly suitable for implementations over quantum information processors (QIPs) with limited $T_1/T_2$ coherence times.  Our shortest circuit has the two-qubit gate depth of only 25,333 (\tab{results})\new{, well within the limits of best trapped ions coherence time of 10 minutes \cite{ar:wuz} divided by the two-qubit gate operating time of about 250 microseconds \cite{comparison}.  The fidelity of physical-level two-qubit gates may not be required to be as low as a simple formula, $\frac{1}{\#\text{gates}}$, prescribes, before a computation with this many gates can be executed.  Indeed, in cases when the dominant error source is of the random under-/over-rotation kind, the error in the execution of the entire circuit may scale sublinearly as a function of the number of gates, and potentially as low as the square root of the number of gates (this is in contrast to additive errors such as those resulting from the decoherence), and thus we can expect to be able to execute a larger number of gates with a given two-qubit gate fidelity.  Sublinear (in the number of gates) scaling of the accumulated error in a quantum computation has already been witnessed experimentally \cite{comparison}, despite small circuit sizes considered.  Applying the mixing unitaries approach \cite{MixedUnitaries} to the physical-level circuits through drawing a pulse sequence from a probability distribution given by multiple approximations that mix to a high-quality two-qubit gate can further lower the requirements for the two-qubit gate error necessary to carry out the computation we proposed.  The lower bound on the gate fidelity of $\frac{1}{\sqrt{\#\text{gates}}}$, being about $0.001$ for $\#\text{gates}{=}$1,000,000 was already achieved experimentally \cite{ar:gtl}, suggesting that the two-qubit gate fidelity required to carry out the proposed computations may be within the reach. Finally, the number of qubits available in either superconducting circuits (72) or trapped ions (79) technology readily meets the qubit count requirement in our circuits \cite{LargeQC}.}

We believe circuit implementations reported in this paper could be particularly relevant to the near-term QIPs based on the trapped ions technology.  This is because the low-diameter graphs, suitable for the kinds of experiments we considered, have connectivities that require long-range interactions between qubits. \new{For 50 or larger number of qubits, a trapped-ion quantum simulator is already known to be capable of all-to-all coupling \cite{QSIM_ion1,QSIM_ion2} and there already exist a suite systematic methods to generate a control signal that implements an arbitrarily-connected two-qubit gate (see, for instance, \cite{FM})}.  Leveraging all-to-all connectivity of the trapped ions QIP, we expect no overhead cost in shuttling quantum information around, a serious point to consider, as has been pointed out in \cite{comparison}.

\section{Discussion} 

In this paper, we synthesized short quantum circuits that aim to solve a scientifically interesting problem.  Specifically, we considered the Heisenberg Hamiltonian simulation with a random disorder on a suite of graphs with small diameter.  Compared to the previous state of the art, our work shows significant gate count savings, a short circuit depth, while natively relying on the qubit-to-qubit connectivity suitable for the implementation over a reconfigurable quantum computer, such as the trapped ions one.  Specifically, we reported a circuit for simulating Hamiltonian dynamics over $(3{-}5{-}70)$ graph for the time $t=10$ and accurate to within the error $\varepsilon=0.001$ using at most 648,885 $\cnotgate$ gates in depth 25,333 in a pre-fault tolerant implementation and 6,751,395 $\tgate$ gates in a fault-tolerant implementation.  We believe the problem instance we considered is intractable for classical computers and yet the quantum resource estimates are very low, showing the promise for solving interesting problems by a quantum computer in a not-too-distant future.


\section{Methods}
\subsection{Problem}

We consider the problem of simulating Hamiltonian dynamics for time $t$, accurate to within the error $\varepsilon$, where the target Hamiltonian $H$ is defined as follows: 
\begin{align}
H :=  \sum_{(i,j)\in E(G)} (\vec\sigma_x^i \vec\sigma_x^j + \vec\sigma_y^i \vec\sigma_y^j + \vec\sigma_z^i \vec\sigma_z^j) + \sum_{i=1}^{n}  d_i \vec\sigma_z^i,
	\label{eq:heisenberg}
\end{align}
where $\vec\sigma_x^i, \vec\sigma_y^i$, and $\vec\sigma_z^i$ denote Pauli $x$, $y$, and $z$ matrices acting on the qubit $i$, $d_i \in [-1,1]$ are chosen uniformly at random, $G$ is a graph describing the two-qubit interactions, $E(G)$ is the set of its edges, and $n$ is the number of qubits.  Such Hamiltonian is known as the Heisenberg Hamiltonian over a graph $G$ with a random disorder in the $Z$ direction.  It has been studied in \cite{54, 57, 58} in the context of many-body localization.

We chose $n$ to be in the range $50$ to $100$.  This is because the largest quantum circuit (vector state as opposed to full unitary) simulations demonstrated to date were achieved with dozens of qubits over a circuit depth of about $40$ \cite{arXiv:1805.01450, arXiv:1804.04797}.  This means that with anywhere more than $50$ qubits and depth in excess of, say, $200$, the problem of quantum circuit simulation may likely become intractable for a classical computer with the simulation techniques such as \cite{arXiv:1805.01450, arXiv:1804.04797}.  Note that our circuits, despite numerous optimizations applied, remain substantively deeper than those considered in \cite{arXiv:1805.01450, arXiv:1804.04797}, motivating our choice to consider reducing the number of qubits from as many as $144$ for a shallow circuit with depth $27$ \cite{arXiv:1805.01450} down to $50$ at the cost of significantly extending the anticipated length of the computation.  We also note that the underlying qubit-to-qubit connectivity pattern in our circuits does not appear to allow breaking the qubit interaction graph into two components by cutting a small number of edges, which lies at the core of efficient simulations such as \cite{arXiv:1805.01450, arXiv:1804.04797}.  The smallest number of qubits we chose to consider, $50$, exceeds the number $42$ used in the best high-depth state-vector type simulation \cite{arXiv:1601.07195}.  Finally, we highlight that the largest numerical simulation of the kind of Hamiltonian we consider is restricted to just $22$ qubits \cite{54}, whereas our work focuses on the Hamiltonians over at least $50$ qubits.

We chose the evolution time $t=2d$, where $d$ is the diameter of graph $G$, similarly to \cite{arXiv:1711.10980}.  The motivation behind such choice is as follows: it takes time at most $d$ \cite{ar:lr} for quantum information to propagate from any node in the underlying graph $G$ to any other node, as such, one may expect to enter a highly entangled simulation regime by the simulation time of $d$.  This means that our simulation spends at least half the time evolving in the regime that we believe is difficult to simulate classically.  Note that due to the use of the product formula approach \cite{Suzuki, 11} in our simulations, circuits for other selection of time $t$ can be developed as effortlessly as changing the number of times a certain block operation is applied. 

A previous study \cite{arXiv:1711.10980} showed that the product formula algorithm(s) \cite{Suzuki, 11} for Hamiltonian simulation with a heuristic bound yields best practical results.  We note that choosing a small diameter graph $G$, such as what we do next, is natural, given the discussion in the previous paragraph.  However, small diameter graphs require large number of edges, and the number of edges in the graph $G$ is directly proportional to the circuit complexity of the single stage of product formula.  Thus, the balance between graph diameter and the number of edges has to be chosen carefully so as to minimize quantum computational resources while maximizing the expected classical difficulty of the simulation. 

We chose graph $G$ to be a minimal distance $k$-regular graph over $n$ nodes.  We select $k$ such that the effort required to develop the individual two-qubit gates for all $\frac{nk}{2}$ interactions prescribed by the graph in a technology such as the trapped ions is not large \cite{PulseShape}, while keeping the graph distance small (resulting in a short enough time $t$ for the evolution before Hamiltonian dynamics simulation enters a regime that is believed to be classically difficult) and the number of qubits as large as possible (while keeping it to between $50$ and $100$).  In practice, we selected the following values of $k$: $3, 4, 5, 6,$ and $7$.  The specific regular graphs $G$ considered in our work can be described by the respective degree-diameter-nodes triple $(k{-}d{-}n)$, as follows: $(3{-}5{-}70)$ Alegre-Fiol-Yebra graph \cite{ar:afy}, $(4{-}4{-}98)$ graph by Exoo \cite{www:e}, $(5{-}3{-}72)$ graph by Exoo \cite{www:e}, and $(7{-}2{-}50)$ Hoffman-Singleton graph \cite{ar:hs}.  The largest known 6-regular distance-2 graph has $32$ nodes (less than $50$ targeted in our work), and the largest known 6-regular distance-3 graph has $110$ nodes (more than $100$ targeted in this work) \cite{www:w}.  Therefore, we did not consider 6-regular graphs.  Note that our goal was to select a graph with a large number of nodes, small degree, and small distance.  A plenty of such graphs can be developed so long as one chooses degree and distance parameters above the minimums known for a given $n$, selects $n$ smaller than the known maximum for the fixed degree and diameter, and expands the attention to graphs other than the regular kind.  Our selection of graphs is very restrictive so as to narrow down the set of specific graphs explicitly considered in our work to a few.  We believe the performance of the Hamiltonian simulation over similar graphs to those considered can be similar.

We chose the approximation error $\varepsilon = 0.001$, measured as the spectral norm distance between the target ideal evolution and the evolution obtained by running our quantum simulation circuits, as in \cite{arXiv:1711.10980}.  This choice of the error value is largely arbitrary, but some choice is necessary to explicitly construct the circuits.

\subsection{\new{Classical hardness}}
\new{We next show that the type of Hamiltonian dynamics simulation problem considered in our work is difficult for classical computers under the assumption that Polynomial Hierarchy does not collapse  \cite{ar:aa}.  Thus, it is likely that quantum computers will have to be used to solve sufficiently large instances of the respective problem (e.g., sampling from the distribution given by the evolution of the given Hamiltonian).}

\new{To prove classical hardness of dynamics simulation of the type of Hamiltonian studied in our work (\ref{eq:heisenberg}), consider this Hamiltonian to be an instance drawn from a larger set, $\mathcal{H}$, being the set of Hamiltonians with time-dependent Heisenberg interactions that can be controllably turned on and off, and a random disorder that can apply in any fixed direction for a given qubit, $X$, $Y$, or $Z$, and can also be turned on and off controllably with time.  Each Hamiltonian in the set $\mathcal{H}$ applies qubit-to-qubit interactions as determined by the respective small-degree small-diameter graph it is considered over.}

\new{The specific instance of the Hamiltonian dynamics simulation problem studied in this work constitutes the hardest instance to simulate in the set $\mathcal{H}$ using a quantum digital computer with the algorithms considered.  This is because all two-body interactions are always on and thus, due to the use of Suzuki-Trotter approach, require a maximal number of gates to be simulated; note that when an interaction is turned off, it takes no gates to simulate the evolution for such periods of time, thereby reducing the gate counts.  Similarly, single-qubit disorders take no gates to simulate for those times when they are turned off (thereby resulting in the $\tgate$ count reduction in the fault-tolerant case).  Random disorders pointing in a given fixed direction $X$, $Y$, or $Z$ does not change how various optimizations reduce the relevant gate counts; this is because commuting rotations over a fixed axis always add into a combined rotation, being the property that we base the resource reductions on.  Furthermore, basis changes $\rxgate{\mapsto}\rzgate$ and $\rygate{\mapsto}\rzgate$ used in the circuit decompositions are accomplished via the use of Hadamard and Phase gates, and do not affect any of the costing metrics considered. }

\new{To establish classical hardness \cite{ar:aa}, we show that the Hamiltonians in the set $\mathcal{H}$ can be used to obtain universal evolutions.  Specifically, we construct a computationally universal library consisting of arbitrary single-qubit and the CNOT gates.  To assist with showing computational universality, consider a Hamiltonian such that all disorders but one point in the $Z$ direction, and there is precisely one qubit $q$ with disorder pointing in the $Y$ direction. 
\begin{itemize}
\item{Arbitrary single-qubit gate.} To construct arbitrary single-qubit gate rely on Euler's angle decomposition in the form $\rygate(\alpha)\rzgate(\beta)\rygate(\gamma)$.  First, consider applying arbitrary single-qubit rotation to the qubit $q$. $\rygate$ gate can be induced by directly evolving the disorder while all other disorders/interactions are turned off.  To apply $\rzgate$, use Heisenberg interaction with the maximal rotation angle to $\swapgate$ qubit $q$ into any of the neighboring qubits, apply $\rzgate$ using the disorder, and finally $\swapgate$ back.  To apply an arbitrary single-qubit gate to any other qubit, first $\swapgate$ it into qubit $q$ (note that the length of the $\swapgate$ chain is at most logarithmic due to the choice of the underlying connectivity graph), and apply the above algorithm.
\item{$\cnotgate$.} To construct the $\cnotgate$ gate, use the following circuit, relying on the operations implementable with the Hamiltonian considered.
\begin{eqnarray*}
\Qcircuit @C=1em @R=2.2em {
	&\ctrl{1} & \qw \\
	&\targ & \qw
}
&
\raisebox{-1.4em}{\hspace{1mm}$\equiv$\hspace{2mm}}
&
\Qcircuit @C=0.7em @R=1.1em{
&\qw  					& \gate{\rzgate(\pi/2)}  & \multigate{1}{\text{Heisenberg}(\pi/8)}	& \gate{\rzgate(\pi)}	& \multigate{1}{\text{Heisenberg}(\pi/8)}	& \qw  					& \qw \\
&\gate{\rygate(-\pi/2)} & \gate{\rzgate(-\pi/2)} & \ghost{\text{Heisenberg}(\pi/8)} 		& \qw		     		& \ghost{\text{Heisenberg}(\pi/8)}			& \gate{\rygate(\pi/2)}	& \qw 
}
\end{eqnarray*}
\end{itemize}
}

\new{Note that we proved classical hardness for a Hamiltonian that is slightly different from the one explicitly studied in this work.  Our motivation was to select a Hamiltonian already used in the context of many-body localization \cite{54, 57, 58}, being a potential application of our study.  We stress that the Hamiltonian used to prove classical intractability can be implemented with at most as many gates as the one we studied in our paper in the pre-fault tolerant case, and using fewer gates in the fault-tolerant case.  This is because we can equalize the strengths of the single-qubit disorders and thus implement those with the input weight algorithm discussed later in the paper, leading to additional reductions in the $\tgate$ count. }

\subsection{Solution} 

\subsubsection{\new{Algorithm}}
Suzuki-Trotter formula based algorithms with empirical bound showed themselves as a potent candidate among quantum algorithms that simulate Hamiltonian dynamics \cite{Suzuki, 11,arXiv:1711.10980}.  We therefore chose to focus exclusively on this type of algorithms.  Specifically, for a Hamiltonian of the form $H = \sum_j \alpha_j H_j$, we approximate the evolution operator according to
\begin{equation}
\exp\biggl(-it\sum_{j=1}^{L}\alpha_jH_{j}\biggr) \approx [S_{2k}(\lambda)]^{r},
\end{equation}
where $\lambda := -it/r$ and
\begin{align}
S_{2}(\lambda)&:=\prod_{j=1}^{L}\exp(\alpha_jH_{j}\lambda/2)\prod_{j=L}^{1}\exp(\alpha_jH_{j}\lambda/2), \nonumber \\
S_{2k}(\lambda)&:=[S_{2k-2}(p_{k}\lambda)]^{2}S_{2k-2}((1-4p_{k})\lambda)[S_{2k-2}(p_{k}\lambda)]^{2},
\label{eq:recursive_def}
\end{align}
with $p_{k}:=1/(4-4^{1/(2k-1)})$ for $k>1$ \cite{Suzuki}.

We selected 4th ($k{=}2$) and 6th ($k{=}3$) order formula versions since in our case they achieved the best results. For a given order formula, the algorithm applies the operation $S_4/S_6$ $r$ times, where $S_4/S_6$ contains $10/50$ repetitions of the product of exponentials of the individual terms of the Hamiltonian $H$, and $r$ is the number of repetitions of $S_4/S_6$ for a given selection of the evolution time $t$, desired accuracy $\varepsilon$, underlying graph $G$, and the selection of random disorders $d_i$ \eq{heisenberg}.  We note that for a fixed $\varepsilon$, graph $G$, and a given selection of random disorders, $r$ is proportional to a growing function of $t$.  Explicit bounds on the value $r$ are $O(t^{1+1/4})$ for the 4th order formula and $O(t^{1+1/6})$ for the 6th order formula \cite{11}.  

\subsubsection{\new{Codesign Principle}}
Since the selection of $t$ in our implementation is $t=2d$, it is important to minimize the diameter of the underlying graph $G$, which explains our focus on the low-diameter graphs.  The cost of $S_4/S_6$ can be described as $10X/50X$, where $X$ is the gate cost of the implementation of a single stage of the product of individual terms in the target Hamiltonian.  In our work, we obtained physical-level implementations of the $X$ stage with the cost of $\frac{3nk}{2}$ CNOT gates and the two-qubit depth $3k$ to $3k{+}3$, and fault-tolerant cost of 
$O(nk)$ $\tgate$ gates and $O(nk)$ CNOT gates with depth $O(k(\log(n)+a))$, where parameter $a$ is defined as the depth of the approximation of an $\rzgate$ (see below for details).  This means that all figures of merit are linearly dependent on the degree $k$ of the graph $G$, and therefore to achieve the best performance, the degree $k$ must be minimized.

\subsubsection{\new{Circuit Design}}

To optimize the depth of our circuits, we employed a customized version of the implementation of Vizing's theorem \cite{co:ju}, which guarantees circuit depth $k$ or $k{+}1$ for the implementation of a single stage of the product of exponentials of the target Hamiltonian.  Our modification includes additional heuristic that randomly reorders the list of edges of the graph $G$ in an attempt to find a depth-$k$ layout when a depth $k{+}1$ layout is found, though in practice we found this to be of limited use.  Our modification also readily accepts a manual input in case a depth-$k$ layout is known, although we chose not to employ this method for the circuits we consider in this paper, since in general the problem of finding a depth-$k$ layout, if at all exists, is NP-complete \cite{EdgeColor} and thus it is unlikely that the appropriate manual input would be known for a generic graph.

\begin{figure}[t]
\centering
\adjincludegraphics[Clip={0.1\width} {0.86\height} {0.\width} {0.06\height},scale=1]{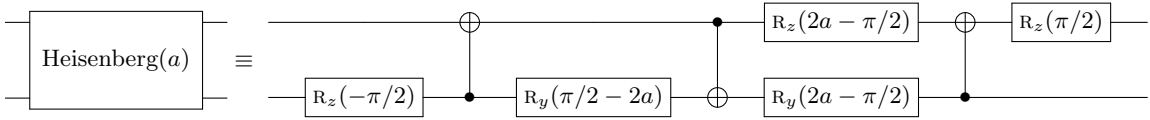}
\caption{$\cnotgate$ count optimal implementation of the Heisenberg interaction up to a global phase of $e^{-i\pi/4}$. This is a special case of the circuit shown in Fig.~6 of \cite{ar:VW}. Note that this circuit requires 3 $\cnotgate$ gates.}
\label{fig:Heisen_pFT}
\end{figure}

\begin{figure}[t]
\centering
\adjincludegraphics[Clip={0.1\width} {0.86\height} {0.\width} {0.06\height},scale=1]{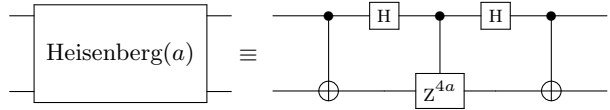}
  \caption{Implementation of the Heisenberg interaction, optimal in the number of real-valued degrees of freedom, up to a global phase of $e^{-ia}$.  Implementation of the controlled-$\zgate^a$ can be substituted from \fig{FTCZ}.}
  \label{fig:Heisen}
\end{figure}

\begin{figure*}[t]
\adjincludegraphics[Clip={0.1\width} {0.75\height} {0.\width} {0.06\height},scale=1]{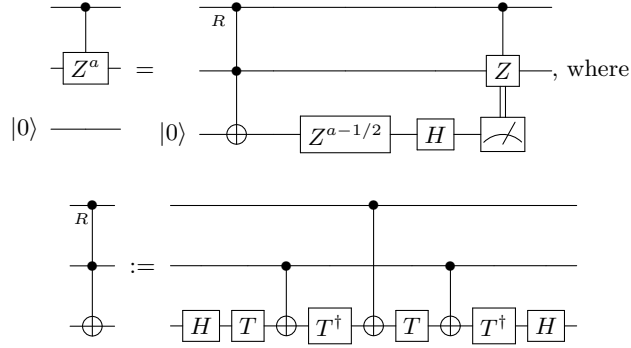}
\caption{Ancilla-aided, measurement/feedforward-based fault-tolerant controlled-$\zgate^a$ gate, imported from \cite{optQFT}.}
\label{fig:FTCZ}
\end{figure*}

We implemented the Heisenberg interaction, $\exp[-ia(\vec\sigma_x^i \vec\sigma_x^j + \vec\sigma_y^i \vec\sigma_y^j + \vec\sigma_z^i \vec\sigma_z^j)]$, using two different circuits, depending on whether we focus on saving quantum resources in a pre-fault tolerant (physical-level) or a fault-tolerant implementation.  In particular, for the physical-level implementation, we used the circuit shown in \fig{Heisen_pFT} in order to minimize the most expensive two-qubit gates, whereas for the fault-tolerant implementation, we used the circuit shown in \fig{Heisen} in order to minimize the most expensive $\tgate$ gates. By directly synthesizing the Heisenberg interaction, compared to the standard Pauli-matrix basis approach, for instance employed in \cite{arXiv:1711.10980}, we save $50\%$ of the cost in the pre-fault tolerant implementation (evidenced through the reduction of the $\cnotgate$ gate count from 6 down to 3) and almost $66\%$ of the cost in the fault-tolerant implementation (evidenced through the reduction from 3 $\rzgate$ gates down to 1 $\rzgate$ and 4 $\tgate$ gates). Because our construction directly implements the Heisenberg interaction, as opposed to an approximate implementation with XX, YY, and ZZ interactions that arise from the standard approach, our implementation also performs better at the algorithmic level, {\em i.e.}, we do not need as large value of $r$ as in the standard approach to keep the overall error level down that incurs from the approximate Pauli-basis implementation.

We laid out the Heisenberg interaction terms in the circuit implementation of the product formula algorithm in alternate orders---forward and reverse---to ensure we obtain maximal gain from the  application of the circuit optimizer \cite{Optimizer}.  We furthermore modified the original optimizer \cite{Optimizer} to ensure it can natively handle controlled-$\zgate^a=\czgate^a$ gates and apply the merging rule $\czgate^a(x,y)\czgate^b(x,y) \mapsto \czgate^{a+b}(x,y)$.  Also implemented was the merging of two Heisenberg interactions as per circuit implementation in \fig{Heisen_pFT} as an optimization rule, and the capability of being able to handle classically controlled quantum gates.  The quality of optimization by the automated optimizer ranged from $7\%$ to $14\%$ in the $\cnotgate$ gate count reduction, and $16\%$ to $24\%$ in the $\rzgate$ count reduction, with the simulations over lower degree graphs yielding a better quality of optimization.  Indeed, circuit implementations over graphs with lower degree have a larger proportion of gates at the edges of the circuit implementation of the Hamiltonian terms, and these are the types of gates that often admit optimizations by techniques such as \cite{Optimizer}. 

\subsubsection{\new{Fault-tolerant Circuits}}
For fault-tolerant implementations, we decided to break the error budget evenly between the algorithmic errors that arise from the product formula algorithm and the approximation errors that arise from the approximation of $\rzgate$ gates in the Clifford$+\tgate$ basis.  We distributed the approximation error budget evenly across all $\rzgate$ gates in the given circuit.  We employed three optimization strategies for fault-tolerant implementations: the mixing unitaries approach detailed in \cite{MixedUnitaries}, the application of equal-angle $\rzgate$ rotations through computing input weight \cite{Gidney}, and a combination of gridsynth \cite{gridsynth1, gridsynth2} and repeat-until-success (RUS) \cite{RUS} strategies for the approximation of $\rzgate$ gates by Clifford$+\tgate$ circuits.  These three strategies are detailed in the next four paragraphs. 

The mixing unitaries approach \cite{MixedUnitaries} relies on generating a set of four approximating circuits for a given $\rzgate(\theta)$ gate, and then applies a randomly drawn approximation from the set of four for every occurrence of the $\rzgate(\theta)$ gate in the circuit, subject to a certain probability distribution.  We found out through simulation experiments that in practice the mixing unitaries approach achieves less than quadratic \cite{MixedUnitaries} improvement in the error.  This is because we measure the expected distance between the approximating circuit and the desired evolution, whereas \cite{MixedUnitaries} measures the error $||\sum_k  p_k U_k - V||$ between a collection of approximating circuits $U_k$ given by the probability distribution $\{p_k\}$ and the target evolution $V$.  Clearly, our metric is more restrictive, and in case when the probability distribution by a collection of approximating unitaries is an acceptable metric, further gate count reductions to those we reported in this paper become possible. 

To study practical performance of the mixing strategy, we chose to investigate an amenable sample case of $(5{-}1{-}6)$ graph in depth.  We varied the per-gate error budget by tweaking the power $p$ of $(\varepsilon_{\rm approx.}/(N_{\rzgate}))^{p}$, where $\varepsilon_{\rm approx.}$ is the approximation error budget and $N_{\rzgate}$ is the number of $\rzgate$ gates in the given circuit, since the number of $\tgate$ gates for an approximation sequence scales linearly in the logarithm of the inverse of per-gate error level.  An extensive numerical investigation showed that choosing $p$ between $0.86$ and $0.89$ works well for the direct synthesis method and the values of $p$ between $0.65$ and $0.68$ represent the advantage by the mixing approach.  We therefore chose to use $p = (0.65+0.68)/2 = 0.665$ as the per-gate approximation power for the mixing unitaries method, for all cases we considered.  We believe this is a safe assumption since in the larger circuits the mixing strategy is expected to give better results as it takes more time to properly average out the errors.  Employing the mixing unitaries strategy allows an estimated $\tgate$ count savings of $25\%$ to $33\%$, depending on whether the power $p=0.875$ or $p=1$ is considered as the starting point. 

In our implementation of the two-qubit Hamiltonian interaction, we lay out the respective circuitry in parallel with the help of Vizing's theorem.  This results in the parallel application of as many as $m{\leq}\frac{n}{2}$ $\rzgate(\theta)$ gates with equal rotation angles [Note that the same idea can be applied to $\rzgate$ gates representing random disorders and looking for rotations that are approximately equal or approximately equal up to a multiplication by a small integer power of 2.  We did not yet pursue such optimization since we believe the improvements will be small (by a few percent).].  Such transformation can be accomplished directly at the cost of $m{\cdot}\text{Cost}(\rzgate(\theta))$ $\tgate$ gates, but a better approach is to induce this set of gates via the calculation of the input weight sum and the application of $\rzgate(2^{\lfloor\log(m)\rfloor}\theta)$, $\rzgate(2^{\lfloor\log(m)\rfloor-1}\theta)$, ..., $\rzgate(2\theta)$, and $\rzgate(\theta)$, to the binary digits (qubits) of the integer number ($\lfloor\log(m)\rfloor{+}1$-qubit ket) representing the input weight of the $m$ qubits needing the application of $\rzgate(\theta)$ gates, at the cost of $4(m-\text{Weight}(m)) + (\lfloor\log(m)\rfloor{+}1){\cdot}\text{Cost}(\rzgate(\theta))$ (where $\text{Weight}(m)$ computes the number of ones in the binary expansion of the integer number $m$) $\tgate$ gates.  Indeed, for the implementations we considered $\text{Cost}(\rzgate(\theta))\approx 50$ $\tgate$ gates, and thus the saving in the $\tgate$ count is substantial.  To induce the input weight calculation at the cost of at most $4(m-\text{Weight}(m))$ $\tgate$ gates, we use $m{-}\text{Weight}(m)$ full and half adders.  We perform all lower digit summations first, using full adders as much as possible (it is always possible when there are at least three bits to be added) and while doing so in parallel.  Since both half and full adder need one relative phase Toffoli gate to be computed, costing $4$ $\tgate$ gates each, and only Clifford gates and one measurement to be uncomputed \cite{Gidney}, the $\tgate$ count in the calculation of the input weight and proper reset of ancillae is precisely $4(m-\text{Weight}(m))$.  The overall optimization of the $\tgate$ count from applying parallel $\rzgate$ gates through the calculation of the input weight ranges from $51\%$ to $60\%$, being better for simulations of higher degree graphs.  Indeed, for higher degree graphs, the product formula algorithm expels a higher fraction of resources on implementing Hamiltonian interactions compared to the resources spent on implementing the random disorder.

We use gridsynth \cite{gridsynth1, gridsynth2} to synthesize optimal single-qubit Clifford$+\tgate$ circuits implementing individual $\rzgate$ gates.  A better strategy relies on the RUS circuits \cite{RUS} that use ancilla, measurement, and feedforward to reduce the $\tgate$ count in the implementation of a single $\rzgate$ gate by a factor of $2.5$ on average.  However, a fully automated implementation of the RUS strategy is unavailable, and the implementation we have is labor-intensive.  Thus, we do not compute RUS circuits explicitly, but rather report the respective expected gate counts.  The RUS implementation can be easily included into our software as an external and independent package.  The result of the overall optimization of the Hamiltonian dynamics simulation circuits by applying the RUS approach ranges between $44\%$ and $50\%$, with optimization quality favoring smaller order graphs.  Indeed, those have smaller parts composed with half and full adders that are not optimized by the RUS. 

We note that optimizations described in the last two paragraphs reduce the $\tgate$ count at the cost of introducing a number of $\cnotgate$ gates.  We keep track of the $\cnotgate$ gates to make sure their cost does not overwhelm that of the $\tgate$ gates.

\subsubsection{\new{Empirical Bound}}
To determine the empirical bound on the number $r$ of iterations of the product formula for our problem that aims to address the graphs $(3{-}5{-}70)$, $(4{-}4{-}98)$, $(5{-}3{-}72)$, and $(7{-}2{-}50)$, we considered random regular graphs with $k=3,4,5$, and $7$, generated using random matching approach \cite{ar:w}, where the number of vertices $n$ ranges from $k{+}1$ to $12$.  For the cases when $nk$ is odd and it is impossible to generate the corresponding random $k$-regular graph (the number of edge ends can only be even), we take a degree-$k$ random graph with $n{+}1$ vertices, and remove a randomly selected vertex as well as all edges leading to it.  We then insert edges connecting $\left\lfloor{k/2}\right\rfloor$ non-overlapping pairs of the resulting $k$ degree-$(k{-}1)$ vertices, chosen at random, provided that the introduction of the new edges does not lead to a multi-edged graph. 

\begin{figure}[t]
\adjincludegraphics[Clip={0.1\width} {0.72\height} {0.\width} {0.06\height},scale=1]{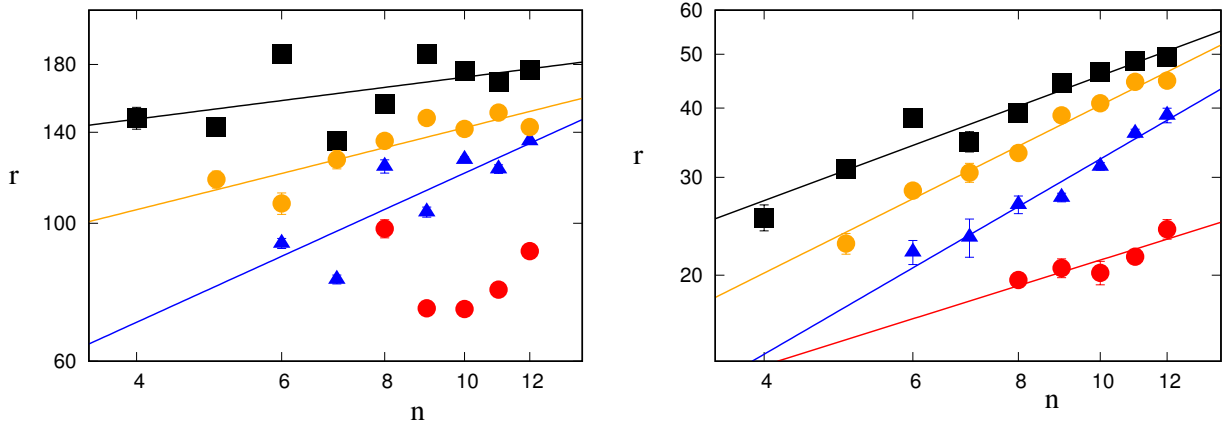}
   \caption{Parameter $r$ scaling data for pre-fault tolerant implementation as a function of the system size $n$ for 4th (left) and 6th (right) order formulas. The black squares, orange circles, \old{and }blue triangles, \new{and red circles} denote $k=3$, 4, \old{and} 5\new{, and} 7 graphs, respectively. The error bars denote one standard deviation. The solid lines are the best fit power law curves (\ref{4th_eq_pFT}) and (\ref{6th_eq_pFT}) for 4th and 6th orders, respectively.}
   \label{fig:rscale_pFT}
\end{figure}

\begin{figure}[t]
\adjincludegraphics[Clip={0.1\width} {0.72\height} {0.\width} {0.06\height},scale=1]{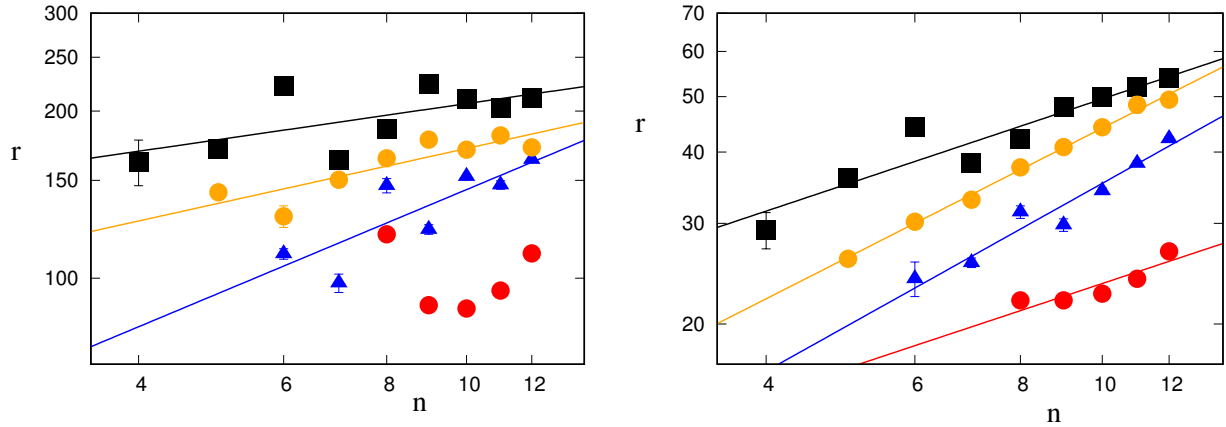}
   \caption{Parameter $r$ scaling data for fault-tolerant implementation as a function of the system size $n$ for 4th (left) and 6th (right) order formulas. The black squares, orange circles, \old{and }blue triangles\new{, and red circles} denote $k=3$, 4, \old{and} 5\new{, and} 7 graphs, respectively. The error bars denote one standard deviation. The solid lines are the best fit power law curves (\ref{4th_eq_FT}) and (\ref{6th_eq_FT}) for 4th and 6th orders, respectively.}
   \label{fig:rscale_FT}
\end{figure}

\fig{rscale_pFT} and \fig{rscale_FT} show, for pre-fault tolerant and fault-tolerant implementations, respectively, the $r$-scaling for the 4th and 6th order formulas for degree $k=3,4,5,$ and $7$ random regular graphs.  By performing least square linear fitting on the log-log scale we determined the following scaling of $r$ in the pre-fault tolerant case, 4th order formula
\begin{equation}
\label{4th_eq_pFT}
r_{3}{=}116.3 n^{0.169}, \; r_{4}{=}66.4 n^{0.331},\; r_{5}{=}30.1 n^{0.602},
\end{equation}
in the pre-fault tolerant case, 6th order formula
\begin{equation}
\label{6th_eq_pFT}
r_{3}{=}12.5 n^{0.564}, \; r_{4}{=}7.05 n^{0.759},\; r_{5}{=}4.24 n^{0.883}, \; r_{7}{=}7.11 n^{0.476},
\end{equation}
while for the fault-tolerant case, 4th order formula
\begin{equation}
\label{4th_eq_FT}
r_{3}{=}126 n^{0.215}, \; r_{4}{=}80.5 n^{0.328},\; r_{5}{=}34.6 n^{0.620},
\end{equation}
and for the fault-tolerant case, 6th order formula
\begin{equation}
\label{6th_eq_FT}
r_{3}{=}15.9 n^{0.493}, \; r_{4}{=}7.82 n^{0.751},\; r_{5}{=}5.26 n^{0.826}, \; r_{7}{=}7.61 n^{0.490},
\end{equation}
where $r_k$ is the empirical bound for the degree-$k$ graph with evolution time $t = 2d =10,8,6,4$ for graphs with $k=3,4,5,7$, respectively.  We did not include the scaling for the 4th order formula and $k{=}7$, since there were too few points to study and there did not seem to be enough stability in the data. \new{However, we investigated different values $\epsilon$ and empirically confirmed the stability in the scalings for $k=3,4,5$.}

Detailed results showing concrete gate counts for individual cases are available in \tab{gc4} and \tab{gc6} for the 4th and 6th order formulas, respectively. Also shown are the gate counts expected for RUS approach \cite{RUS}, where, for each $\rzgate$ approximation, we expect a factor $2.5$ reduction in the $\tgate$ count at the cost of the addition of at most ($\tgate$-count$+1$) $\cnotgate$ gates and one ancilla. 



\begin{acknowledgments}
Authors thank \new{Prof. Scott Aaronson (University of Texas -- Austin) and} Prof. N. Julien Ross (Dalhousie University) for helpful discussions. 

This material was based on work supported by the National Science Foundation, while DM working at the Foundation.  Any opinion, finding, and conclusions or recommendations expressed in this material are those of the author and do not necessarily reflect the views of the National Science Foundation.
\end{acknowledgments}




\begin{table}[ht]
\scriptsize
\renewcommand\thetable{2}
  \caption{Gate counts for different graphs using 4th order PF}
  \label{tab:gc4}
  \tabletwo
\end{table}

\begin{table}[ht] 
\scriptsize
\renewcommand\thetable{3}
  \caption{Gate counts for different graphs using 6th order PF}
  \label{tab:gc6}
  \tablethree
\end{table}

\end{document}